\documentclass{article}  

\usepackage{mathtools}
\usepackage{amsfonts}
\usepackage{algorithm}
\usepackage[noend]{algpseudocode}
\usepackage{graphicx}
\usepackage{wrapfig}
\usepackage{dblfloatfix}
\usepackage{url}
\usepackage{color}
\usepackage[font=small,labelfont=bf]{caption}
\usepackage{subcaption}
\usepackage{graphicx}
\usepackage{lipsum}
\usepackage{soul}

\usepackage{float}

\DeclareCaptionFont{tiny}{\tiny}

\title{\LARGE \bf
Investigating usability of MSstatsQC software
}

\author{Sara Mohammad Taheri, Omkar Terse, Eralp Dogu, and Olga Vitek
}

\begin{document}

\maketitle
\thispagestyle{empty}
\pagestyle{empty}

\begin{abstract}

MSstatsQC \cite{dogu2017msstatsqc} is an open-source software that provides longitudinal system suitability monitoring tools in the form of control charts for proteomic experiments. It includes simultaneous tools for mean and dispersion of suitability metrics and presents alternative methods of monitoring through different tabs that are designed in the interface. This research focuses on investigating the usability of MSstatsQC software and the interpretability of the designed plots.

In this study, we ask 4 test users, from the proteomics field, to complete a series of tasks and questionnaires. The tasks are designed to test the usability of the software in terms of importing data files, selecting appropriate metrics, guide set, and peptides, and finally creating decision rules (tasks 1 and 3 in appendix). The questionnaires ask about  interpretability of the plots including control charts, box plots, heat maps, river plots, and radar plots (task 1 and 4 in appendix). The goal of the questions is to determine if the test users understand the plots and can interpret them.

Results show limitations in usability and plot interpretability, especially in the data import section. We suggest the following modifications. I) providing conspicuous guides close to the window related to up-loading a data file as well as providing error messages that pop-up when the data set has a wrong format II) providing plot descriptions, hints to interpret plots, plot titles and appropriate axis labels, and, III) Numbering tabs to show the flow of procedures in the software
\end{abstract}

\maketitle

\section{Introduction}

Mass spectrometry proteomics is a  technology that is used to quantitatively study the protein compounds of complex biological organisms. These experiments require several stages, including sample preparation, separation through liquid chromatography, and measurement through mass spectrometry. Although there are technologies that are developed to help improve the replicability and reproducibility of the results, variability exists and is an issue. This variablity have several sources. Therefore, automated longitudinal monitoring of system suitability and quality control (QC) is essential to ensure replicability and reproducibility of quantitative proteomic workflows. \cite{dogu2018msstatsqc}.

In laboratories there are experiments that run for an extended time and have many replicates. It is very important to ensure the measurement system for the experiments performs as intended. Reasons for a measurement system's deviation from expected performance include different laboratories, instruments, operators, temperature of the room, change in person doing the experiment, or different time of data acquisition. MSstatsQC \cite{dogu2017msstatsqc} is an open-source software that diagnoses these changes early in time and enables intervention. It provides longitudinal system suitability monitoring tools in the form of control charts for proteomic experiments. These tools includes simultaneous control charts for mean and dispersion of suitability metrics and presents alternative methods of monitoring through different plots that are designed in this interface. 

The value of using MSstatsQC is that if there is a deviation from the expected values of the experiment, those experiments can be stopped early in time. This save the cost and time. It also provides an environment for scientists who do not have the required programming skills to perform a computer-based analysis of their results.

There are five main steps to use MSstatsQC. The first is to upload a data set with a required format. Second, the experimenter selects the metrics, determines the guide set to estimate metric mean and variance, and selects specific precursor(s) or select all. The third step is to design appropriate decision rules for when a system performance is poor and unacceptable. The fourth step is to run and generate control charts. The last step is to check decision maps, metric summary plots, and change point analysis for better statistical reasoning.

Issues may arise if an experimenter fail to appropriately do any of the aforementioned steps above. If an experimenter does not use a data set with the correct format, he/she cannot select the metrics in the second step which results in failure in steps three and five. The flow of the steps is important and have dependencies. for example, first the decision rules must be set and then the results will appear in the specific tab for the decision maps. An experimenter must be able to interpret control charts, river plots, radar plots and decision maps to notice when the system fails to perform well.

Our study focuses on the usability of MSstatsQC and the interpretability of its plots. Usability is measured by having a number of test users (volunteers) use the software to perform a prespecified set of tasks and evaluating their experience on the basis of some performance parameters and questionnaire. In designing the study, we defined 4 specific tasks. Two tasks focus on measuring the usability of software in terms of importing data sets, selecting appropriate metrics, guide set, and peptide, and creating decision rules. Two other tasks include questions that focus on interpretability of the plots including control charts, box plots, decision maps, river plots, and radar plots.


We conducted 2 pilots tests where we used as a base to find the issues with the defined tasks and the general flow of the study. After the end of second pilot test, we changed some of the questions, make the descriptions more clear and approximated the time that it will take the real test user to complete the tasks.

We hired 4 candidate test users, 2 male and 2 female, to participate in our study. The candidates are researchers from the mass spectrometry proteomic experiments field and have knowledge about quality control tasks. None of the test users have used MSstatsQC software before.

We found that the least rated attribute is convenience with an average rate of 2.7/5. The test users were not comfortable interpreting the plots because of lack of plot explanation, title and axis labels. They found it hard to read the name of peptides in the radar plots because the names were long and mixed together. None of the test users were able to find the issue with the format of the data in task 1, which shows that there is not enough and bold information about the required format of the data in the software. We realized that it is required that an error message pops up if the data doesn't have the correct format.

MSstatsQC, and other similar software in the community of mass spectrometry based proteomic experiments has never been evaluated for their usability and interpretability of the plots. Our study is very unique and can be used as a guide for other researchers in the community to conduct similar tests and evaluation methods for their software.

The community that will benefits most from our usability study is the field of targeted proteomic experiments. For that, we need to understand these experimenters and their goals in using MSstatsQC software. In terms of the HCI community, we believe that the best fit for our project is the design community. The design of this usability study can save as an example and motivation for usability studies of other tools in the field of proteomic studies. Nevertheless, our main goal in this study is to improve the current design of MSstatsQC user interface.

\section{Background}
In this section we introduce two available interfaces that performs quality control for proteomic studies and compare them with MSstatsQC. 

First interface is AutoQC \cite{bereman2016automated}. It is a web based interface for automated data processing that provides quality control workflow for instruments that are used in proteomics experiments. AutoQC works closely with Skyline \cite{maclean2010skyline} and Panorama \cite{sharma2014panorama}.  Skyline is an open-source software for monitoring proteomics experiments and Panorama is an open-source application for targeted mass spectrometry assays. AutoQC works only with data sets that are generated through Skyline and accepts only 8 specific metrics, whereas MSstatsQC does not depend on another software, and accepts any type of metrics without a limitation on the number of metrics. AutoQC only provides control charts whereas MSstatsQC provides control charts as well as statistical plots such as box plots, river plots, radar plots and decision maps. A nice feature of AutoQC is that it can automaticaly updates the plots as data is generated, so an experimenter can stop the experiments when the system is out of control. MSstatsQC cannot do that. Finally, AutoQC is web-based and does not have a Shiny interface like MSstatsQC.

Second interface is SProCoP \cite{bereman2014implementation} which implements control charts and Pareto analysis into the Skyline \cite{maclean2010skyline} software. It provides real time evaluation of the chromatographic performance and mass spectrometric performance via control charts and box plots. Like AutoQC and unlike MSstatsQC, SProCoP depends on Skyline, accepts only a limit of specific metrics, only provides control charts for change in mean, do not provide statistical plots, , and has the ability of real time monitoring of data. Similar to MSstatsQC, SProCoP has a Shiny interface.

\begin{figure*}
    \centering
    \begin{subfigure}[b]{0.49\textwidth}
        \includegraphics[width=\textwidth]{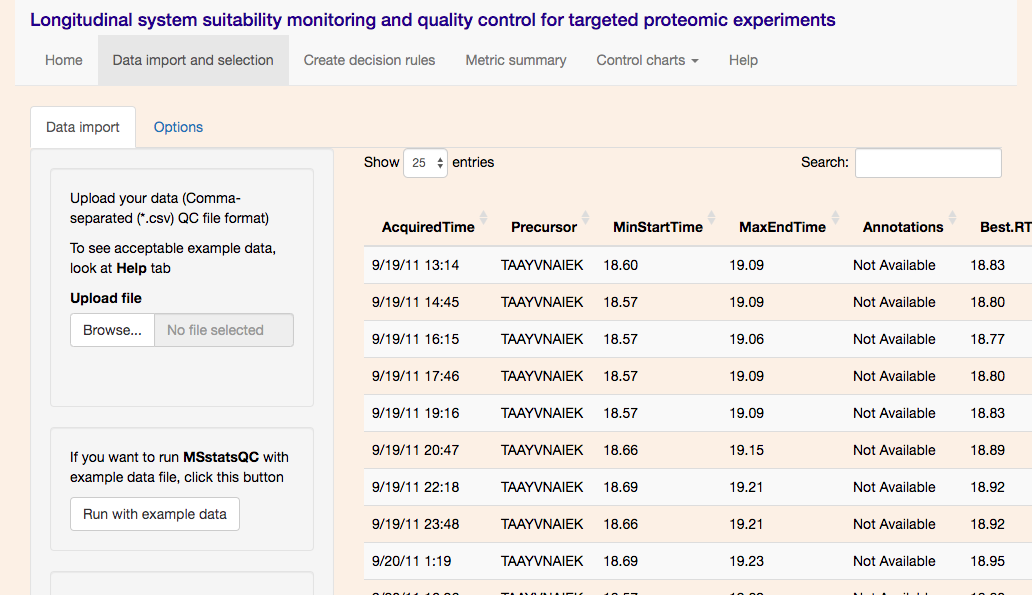}
        \caption{Import the sample data}
        \label{fig:importData}
    \end{subfigure}
    ~ 
    \begin{subfigure}[b]{0.49\textwidth}
        \includegraphics[width=\textwidth]{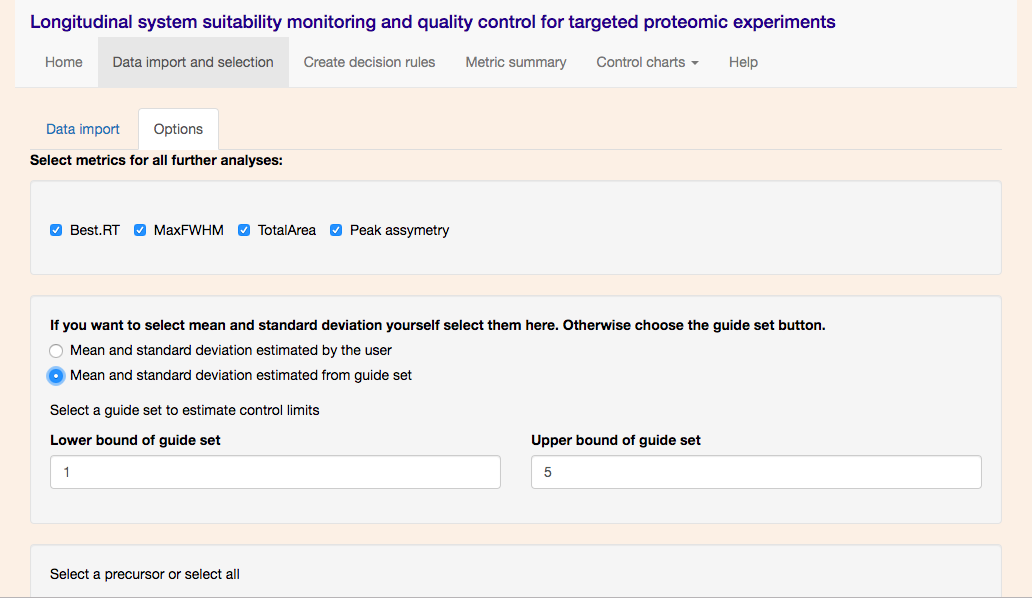}
        \caption{Select metrics, guide set and a specific peptide}
        \label{fig:selectMetric}
    \end{subfigure}
    ~ 
    
    \begin{subfigure}[b]{0.49\textwidth}
        \includegraphics[width=\textwidth]{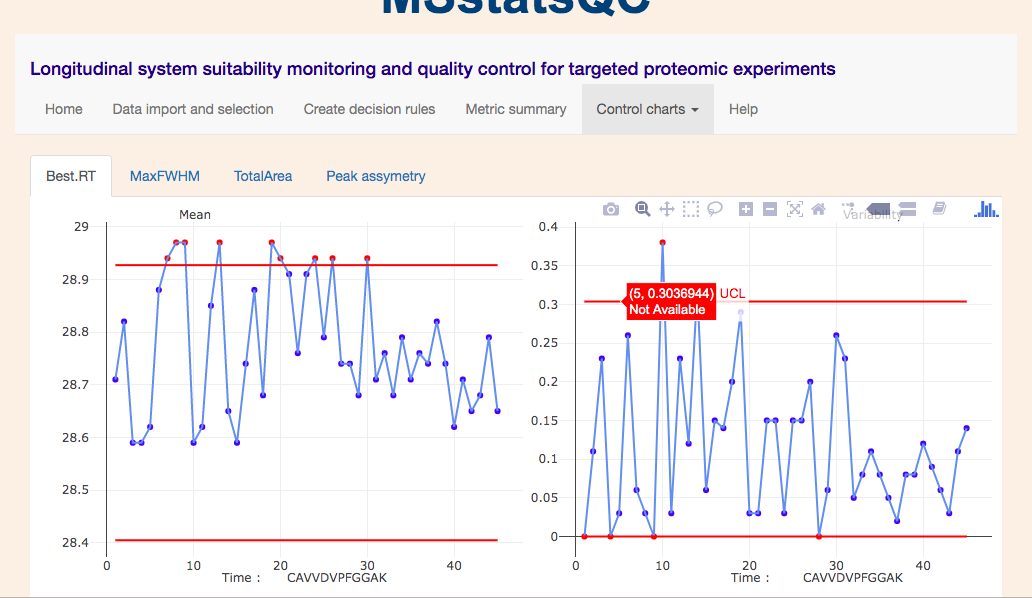}
        \caption{Mean and variability of the best retention time metric}
        \label{fig:controlCharts}
    \end{subfigure}
    \begin{subfigure}[b]{0.49\textwidth}
        \includegraphics[width=\textwidth]{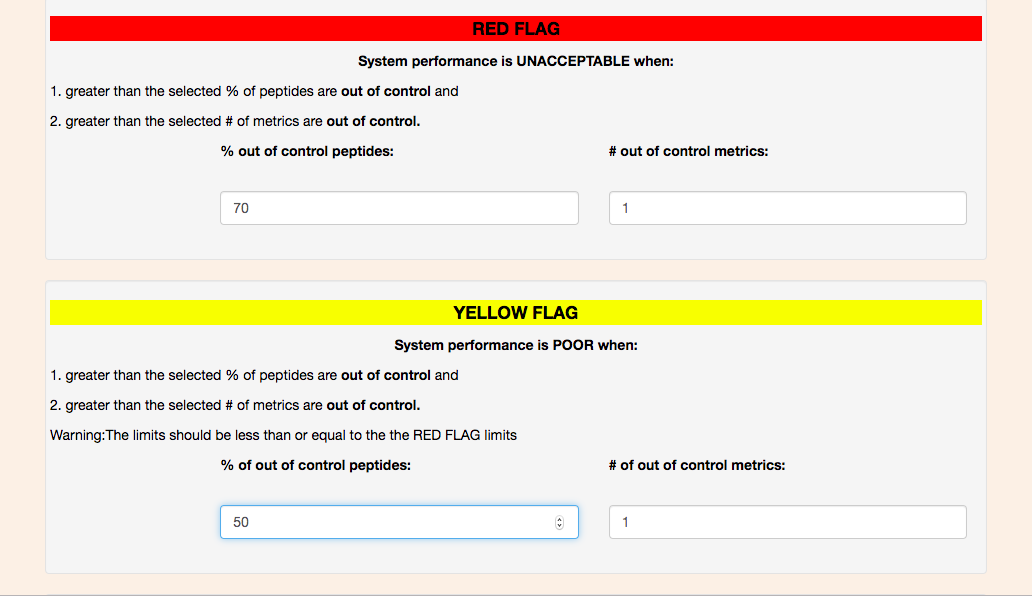}
        \caption{Design performance criteria}
        \label{fig:decisionRules}
    \end{subfigure}
    \begin{subfigure}[b]{0.49\textwidth}
        \includegraphics[width=\textwidth]{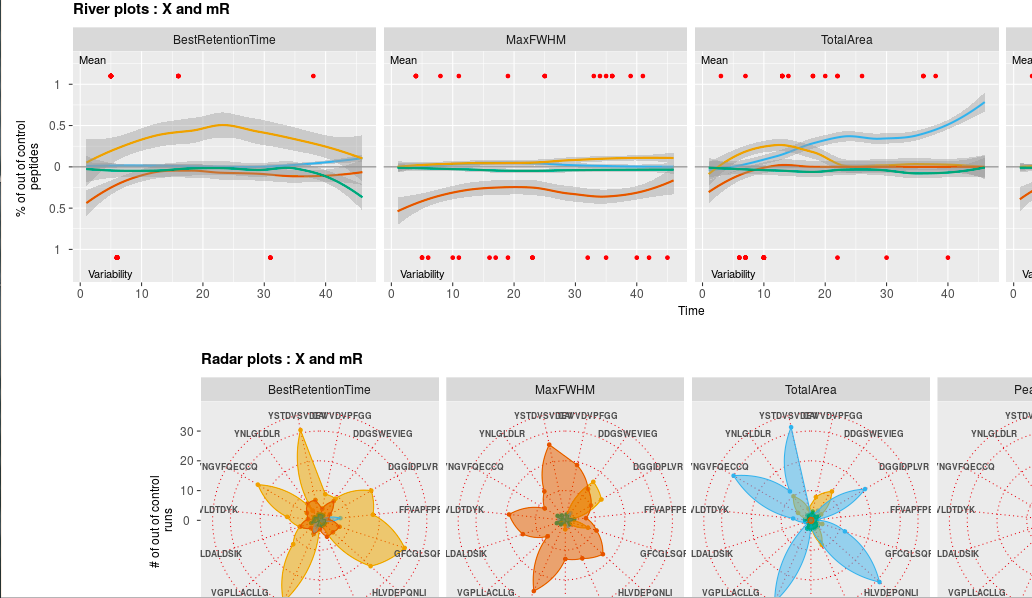}
        \caption{Top: River plots. Bottom: Radar plots}
        \label{fig:riverRadar}
    \end{subfigure}
    
    \begin{subfigure}[b]{0.49\textwidth}
        \includegraphics[width=\textwidth]{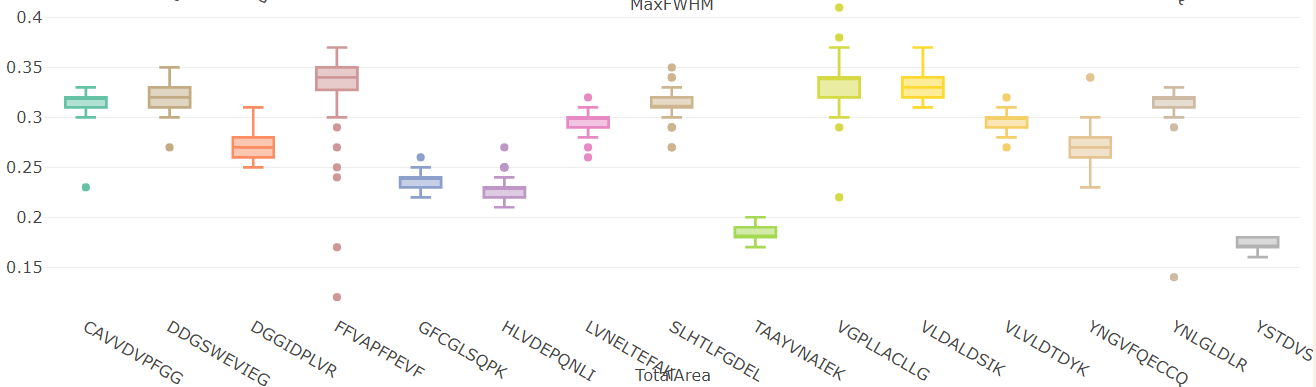}
        \caption{Boxplots}
        \label{fig:summaryBoxplot}
    \end{subfigure}
    \begin{subfigure}[b]{0.49\textwidth}
        \includegraphics[width=\textwidth]{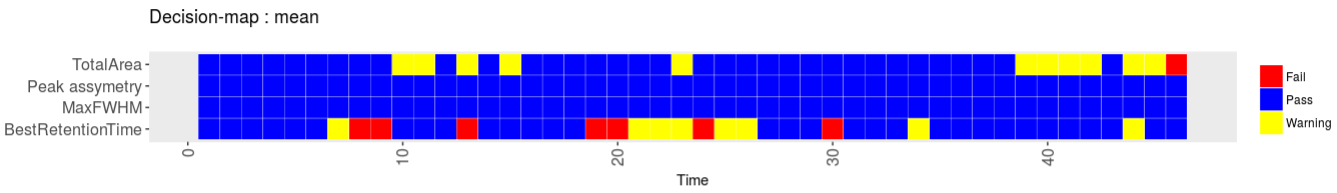}
        \caption{Metric Summary: Decision Maps}
        \label{fig:decisionMaps_mean}
    \end{subfigure}
    \caption{An overview of MSstatsQC. (a) Data Import tab. The experimenter can upload data by clicking on the Upload button. Data set is shown to the left, (b) The experimenter can select metrics, guide set and peptides from the Options tab, (c) Control chart for a metric, (d) Create Decision Rule tab. The experimenter can select specific decision rules for when he/she decides a system performance is poor or unacceptable, (e) Summary Plot tab. It includes river and radar plots.}\label{fig:overviewMSstatsQC}
\end{figure*}


\section{Overview of MSstatsQC software}

MSstatsQC \cite{dogu2017msstatsqc} is an open-source software that provides longitudinal system suitability monitoring tools in the form of control charts for proteomic experiments. It includes simultaneous tools for mean and dispersion of suitability metrics and presents alternative methods of monitoring through different tabs, which are designed in the interface.

The experimenter can upload a data set and analyze the behaviour of the experiment. For example, in Figures \ref{fig:importData} and \ref{fig:selectMetric}, experimenter uploads data, select 4 system suitability metrics, mean and standard deviation of metric, and a specific peptide (CAVV) to monitor the instrument performance. In Figure \ref{fig:controlCharts} experimenter selects a specific metric to review the control charts to find out when the system is out of control. Figure \ref{fig:controlCharts} shows for example, that the mean of best retention time metric for CAVV peptide is out of control for time points 7,8,9,13,19,20,24, 26 and 30 and the experimenter needs to do further analysis to find out the issue with the instrument.

The experimenter is able to design performance criteria by creating decision rules, as in Figure \ref{fig:decisionRules}. Finally, experimenter can view statistical plots such as box plots in Figure \ref{fig:summaryBoxplot}, heat maps in Figure \ref{fig:decisionMaps_mean}, river, and radar plots in Figure \ref{fig:riverRadar}.

\section{Methods}
In this study we recruited 4 test users, who work in field of targeted proteomics experiments. We asked the test users to complete a series of tasks and questionnaires. Two tasks were designed to test the usability of MSstatsQC software in terms of importing data files, selecting appropriate metrics, guide set and peptides, and finally creating decision rules (task 1 and task 3). Two other tasks are mainly based on task 1 and 3, and are a set of questions about the interpretability of the plots including control charts, box plots, heat maps, river plots, and radar plots. The goal of these series of questions is to determine if the test users understand the plots. We introduced each task in more detail in the Results section. They can also be found in the appendix section. Our objective was to gather the following information: 1) time user takes to complete a task, 2) the number of tasks completed within a time limit, 3) frequency of need of instructions, 4) number of errors answering interpretability questions, 5) Ratings of certain features after completion of all the tasks, and 6) number of times the user expresses frustration. We used Google forms to collect our data set from test users. The data set was then converted to an MS-Excel file. Finally we used Tableau software to generate plots to gain some insights about areas of software improvements.\\
\textbf{Time test user takes to complete a task}
Screening records while test users are performing the tasks, we measure the time that it takes for each test user to complete each task. One usability attribute is “learnability”, which says that the “system should be easy to learn so that the experimenter can rapidly do some work done with the software. By measuring the time a test user takes to complete a task, we measure the learnability attribute of usability” \cite{nielsen1994usability}.\\
\textbf{The number of tasks completed within a time limit}
The test users may not be able to complete a task, or they may complete it incorrectly which we will count as an incomplete task. \\
\textbf{Frequency of need of instructions}
We measure this by looking at the screen records to see how often a test user reads the instructions provided for them and how often the test user reads the help page of the software to Figure out how to perform the task. 
\textbf{Number of errors answering interpretability questions}
The tasks that we want to ask from test users are going to measure 2 different areas. One is to find out if the test user finds it convenient to navigate through different tabs in the software and find the steps easily. The other one is to find out if the plots in the software are interpretable and the test user can understand what we want them to understand by looking at different plots. If the test user answers the interpretability questions about the plots incorrectly, it means that the plots are not interpretable enough.\\
\textbf{Ratings of certain features after completion of all the tasks}
After all the tasks are completed we ask the test user to rate some of the features of the software like the interpretability of plots, the ease of usage and the effectiveness of design (colors, design of tabs, etc).\\
\textbf{Number of times the test user expresses frustration}
This involves having the test user  do a given set of tasks while being asked to “think out loud”( cognitive walkthrough). The purpose of this measure is to find out the number of times the test user expresses any signs of frustration. This insight into a user’s thought process can help determine concrete interface elements that cause misunderstanding, so we can redesign them.

\subsection{Pilot tests}
We conducted 2 pilot tests, one user per pilot, with test users that have never worked with MSstatsQC software. The goal of the pilot tests was to make sure that the questions are interpreted properly by the test users. 

The first pilot test revealed a couple of issues in terms of interpretability of the first task, as well as  problem accessing the test data. We identified some of the usability issues with the software during the pilot test, such as lack of an error message in the software when a wrong data set is uploaded. While the first pilot test user was working on the  tasks, we recorded the start and end time for performing each task. 

We ran the pilot test with a second user. After the second pilot test, we corrected the issues with test data set, and revised the questions to ensure  they are interpretable. The second pilot user, did not have any issues with the interpretability of the questions and pointed out some suggestions about how to design some of the plots. We deleted one of the interpretability questions after the second pilot test, because both pilot users indicated a duplication in what different questions elicited. We captured the time in the same way as we did for the first pilot user and made an estimate based on the results that we got from both pilot users. Our estimate of task completion time for all the 4 tasks is around 40 minutes. 
\section{Results and discussion}
In this section, we briefly introduce the 4 tasks that measure the usability of MSstatsQC and interpretability of its plots. 4 test users participated in our study and conducted the tasks. While the test users were performing the tasks, we recorded their audio and screen. We recorded the time it takes each user to complete a task and compared it to our estimated time from pilot tests. After they successfully completed the tasks, we asked test users to rate the system on several attributes of usability. These attributes include user convinience, aesthetics, comprehension of the software and comprehension of peptide name, and satisfaction. All data was recorded through Google Form responses as well as in an MS-Excel spreadsheet. The data (qualitative and qualitative) in the spreadsheet was cleaned, transformed, and visualized using Tableau to gain some insights about areas of software improvements.

\begin{figure*}
    \centering
    \begin{subfigure}[b]{0.3\textwidth}
        \includegraphics[width=\textwidth]{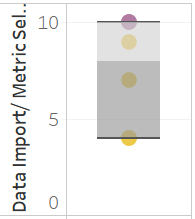}
        \caption{Task 1}
        \label{fig:dataImport}
    \end{subfigure}
    ~ 
    \begin{subfigure}[b]{0.3\textwidth}
        \includegraphics[width=\textwidth]{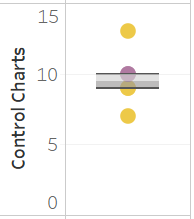}
        \caption{Task 2}
        \label{fig:controlCharts_dist}
    \end{subfigure}
    ~ 
    \begin{subfigure}[b]{0.3\textwidth}
        \includegraphics[width=\textwidth]{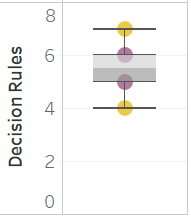}
        \caption{Task 3}
        \label{fig:decisionMaps}
    \end{subfigure}
    
    \begin{subfigure}[b]{0.3\textwidth}
        \includegraphics[width=\textwidth]{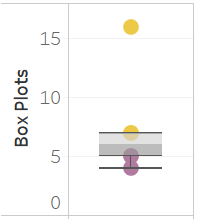}
        \caption{Task 4, part 1}
        \label{fig:summaryBoxplot_dist}
    \end{subfigure}
    \begin{subfigure}[b]{0.3\textwidth}
        \includegraphics[width=\textwidth]{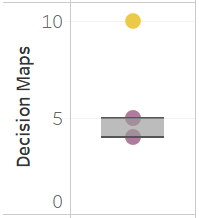}
        \caption{Task 4, part 2}
        \label{fig:decisionmaps_dist}
    \end{subfigure}
    \begin{subfigure}[b]{0.3\textwidth}
        \includegraphics[width=\textwidth]{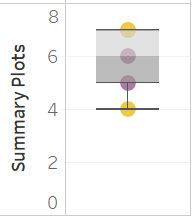}
        \caption{Task 4, part 3}
        \label{fig:summaryplots_dist}
    \end{subfigure}
    \caption{Distribution of the time test user and pilot users take to complete a task. Test users are represented as a yellow point and pilot users as a purple point. (a) Distribution for clocked time for task 1, (b) Distribution of clocked time for task 2, (c) Distribution of clocked time for task 3, (d) Distribution of clocked time for task 4, part 1, (e) Distribution of clocked time for task 4, part 2, (f) Distribution of clocked time for task 4, part 3}\label{fig:performanceTime}
\end{figure*}
\subsection{Results of performing task 1}
The first task for test users was importing a data file, selecting appropriate metrics, guide set, and peptide names. The software requires a specific format for the data set. We asked the test user to upload a data set that doesn't have the correct format, and our goal was to find if the test user can realize that the data does not follow the required format by the software. The other goal of task 1 is to evaluate the usability of the software by measuring how convenient it is for the test user to select appropriate metrics, guide set, and peptide names. 

Figure \ref{fig:dataImport} shows a boxplot that represents the time it takes each user to complete task 1. Test users are represented as a yellow point and pilot users as a purple point. The distribution shows that most test users took less than the estimated time, indicated by the wide inter-quartile range as shown in Figure \ref{fig:dataImport}.

To our surprise, none of the test users were able to successfully complete task 1. They didn't realize that the data set needs to follow a specific format. The main reason for such a failure is because of the lack of a conspicuous guide about the required format. The guide that was provided in the software was written in light gray which did not capture the attention of test users. It was also far from the "upload a data" button. We can improve the usability of software in terms of data import by providing clear instructions about the format of data, make an error pop-up if the data set doesn't have the required format, or as a high-level effort, auto-format the data to remove the data format requirement. Evaluations on usability of the software for metric, peptide and guide set selection showed that the user cannot fluently follow the flow of software. We can improve this issue by numbering different tabs to indicate the flow of the procedure.\\

\subsection{Results of performing task 2}
The second task for test users included a set of questionnaires about interpreting the control charts. The goal of this task is to find whether the user understands the difference between the mean and variation control charts and whether they can detect the abnormalities in an experiment.

Figure \ref{fig:controlCharts_dist} shows a box plot that represents the time it takes each user to complete task 2. The time distribution shows a narrow inter-quartile range with multiple outliers.

All the test users completed this task successfully. However, one user expressed incomprehension while performing this task. Users suggested to us to provide explanations describing the plots and plot features (red/ blue dots, threshold line), provide hints for interpreting the plots, label the Y-axis and provide bold plot titles.


\subsection{Results of performing task 3}
This task was for setting decision rules (shown in \ref{fig:decisionRules}), which will act as thresholds for deciding when a systems' performance is considered as fail and when it is considered as poor. The goal of this task was to find if the concept of setting a user-defined as opposed to a preset-decision rule is clear for the users and if they can set appropriate decision rules.

Figure \ref{fig:decisionMaps} shows a box plot that represents the time it takes each user to complete task 3. The time distribution for this task was uniform. \\

As the quantitative data was not insightful enough, we decided to concentrate on the qualitative data. One great test user suggestion was to miniaturize and combine the Decision Rules tab with the Decision Maps tab as in Figure \ref{fig:decisionMaps_mean}. The reason for this suggestion is that decision rules only affect decision maps, but it is located on a separate tab far from the decision maps tab. It was not clear for the users where after setting the decision rules, they should expect changes. Moving decision rules tab to decision maps shows continuity and eliminates redundant use of interface. Some test users suggested aesthetic improvements, like adding a 'Save' button to indicate to the users that the decision rules are saved and are applied to the Decision Maps. We used blue color to show that a system performance is acceptable, but the users prefer green color, because the combination of green (system performance is acceptable), yellow (system performance is poor), and red (system performance is unacceptable) represents  traffic lights.

\subsection{Results of performing task 4, part 1/3}
The first part of test user task  was about interpreting box plots as a metric summary plot as shown in Figure \ref{fig:summaryBoxplot}. The goal of this task was to find if box plots are a good visualization to show the distribution of peptide values and if the test users are willing to use them for their analysis. Another goal was to find out if the users understand that changing the decision rules does not affect the distribution in box plots.

The time distribution for completing this task shows narrow inter-quartile range with an outlier as in Figure \ref{fig:summaryBoxplot_dist}), indicating test users spend more time in the task than expected. One of of test users did not understand the box plots and we   explained it that user. All test users indicate that they like to use box plots for their analysis. Two of the test users didn't realize that changing the decision rules does not affect the box plot. The reason for this mistake is because the tab that shows box plots comes right after the decision rule tab. As we discussed earlier, we can solve this misunderstanding by moving the decision rules tab to the Decision Plots tab, because decision plots are the only plots that are affected by decision rules. Other areas for improving the interpretability of box plots includes providing explanation for the plots, positioning them better to avoid overlapping of titles, making titles bold,  using first 3 letter of peptide names in the x axis, and highlighting the location of mean in box plots to identify it.

\subsection{Results of performing task 4, part 2/3}
In this task the test users are instructed to interpret the decision maps that are shown in Figure \ref{fig:decisionMaps_mean}. The goal of this task was to find if the test users can interpret the plots and find time points on the plot that the system performance was poor or unacceptable. The other goal was to find if the users understand the difference between the two decision maps where one is based on mean and the other is based on variability. 

The time distribution in Figure \ref{fig:decisionmaps_dist} indicates a narrow inter-quartile range with the presence of a single outlier.

All test users performed the task successfully providing no significant insights for improving the software. Some general suggestions provided by the test users were to provide plot explanations and bold titles.

\subsection{Results of performing task 4, part 3/3}
In this task the test users are instructed to interpret the summary plots that are shown in Figure \ref{fig:summaryplots_dist}. Except for one user, all other clocked timings were present within the inter-quartile range.

\begin{figure}[h]
\caption{Google Form Response Analysis: Correct peptides have are underlined. Bar chart shows that most of the test users have problem in selecting HLV peptide, some of them did not select VLV peptide, and one test user selects other peptide names because the test user didn't understand the question correctly.}
\includegraphics[width=0.5\textwidth]{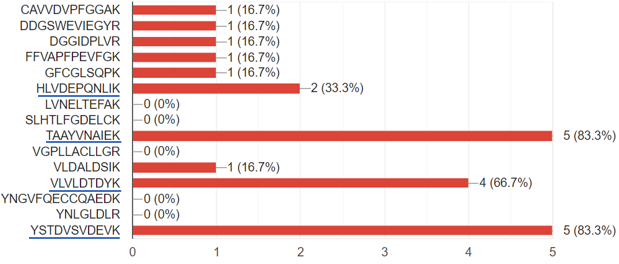}
\label{top-4-pn}
\end{figure}

Figure \ref{top-4-pn} shows the Google Form Response analysis of the test users when we asked them to observe the radar plots for a specific and select the top 4 peptides that have the most metric variability decrease. The underlined peptide response is the correct response, and the plot clearly shows that not all test users chose the correct options. Test users suggested to provide plot explanations, enlarge graphs, provide hints for interpretation, provide different marker symbols to indicate top 3 peptides, and use  first 3 peptide names for labelling.

\subsection{Rating Analysis}
After completing the tasks, test users were asked to rate the system on several attributes of usability. These attributes include user convenience, aesthetics, comprehension of the software, and comprehension of peptide name, and satisfaction.

\begin{figure}[h]
\caption{Weighted Factor Average of Ratings for each test user. The scale is from 1-5. 1 is the lowest rate and 5 is the highest rate.}
\includegraphics[width=0.5\textwidth]{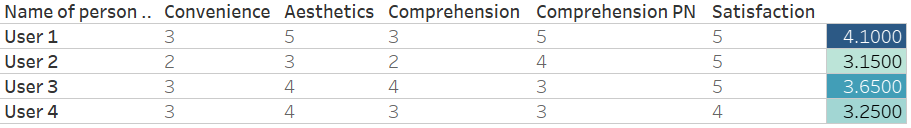}
\label{weightedRating}
\end{figure}

\begin{figure}[h]
\caption{Mean Rating for each attribute}
\includegraphics[width=0.4\textwidth]{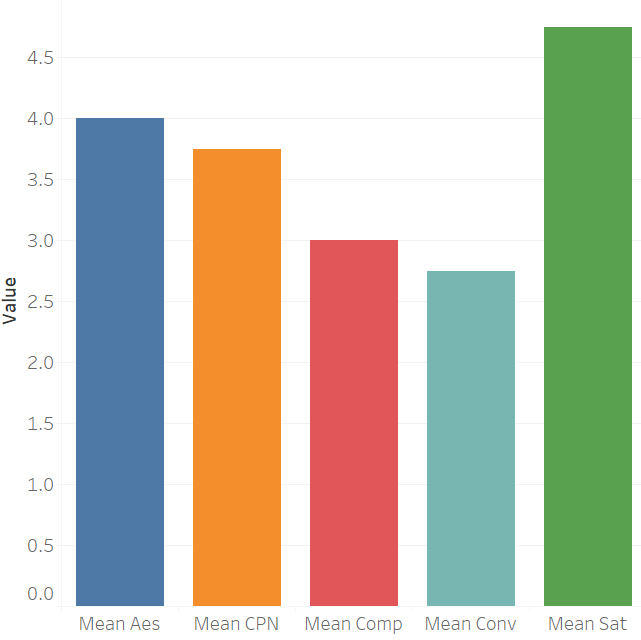}
\label{meanRating}
\end{figure}

We used a weighted factor analysis to get an overall rating (shown in Figure \ref{weightedRating}) and the mean rating (shown in Figure \ref{meanRating}), was calculated for all test results and attributes respectively. For weighted factor average, the weight of 100 was divided for all the attributes depending on the importance of the attribute for the researchers. The distribution of weights is: Comprehension of Peptide Names: 30, System Comprehension: 25, Convenience: 20, Satisfaction: 15, Aesthetics:  10.

The mean rating suggested that the least rated attribute was convenience and comprehension. This low rating is because of lack of enough information and explanation for data import, control charts and summary plots. The plots need bold title, appropriate y-axis label, and 
only represent the first three names of the peptides to avoid messiness in box plots and radar plots.

\section{Limitations and Future Work}
The main limitation of this work was the lack of qualified researchers that can participate as a test user in our study. The community of mass spectrometry proteomics experiments is a small community and researchers ususally are not interested to participate in a 40 minutes study that does not have any reward. The second limitation was the difficulty of designing tasks that cover the usability of the software and interpretability of its plots and are doable within 30-40 minutes. We used our pilot test users to design the tasks appropriately. 

Future work is to redesign MSstatsQC software based on the results that we get in this study, and re-evaluate the new software with another usability test, and show that the attributes of usability such as convenience, satisfaction, comprehension and aesthetics has improved. Another future work would be designing a series of rules and techniques that can serve software designers in the field of proteomics experiments in a way that they can follow the rules to design their own usability studies.

\section{Conclusion}
After reviewing the results of our usability study of the MSstatsQC software, the researchers concluded that the following improvements are necessary and would help increase the usability of the system interface.
The main focus of improvement is convenience. Most of the test users were able to perform most of the designed tasks but they struggled to find which tab they should select to do the next step. There are several tabs and sub tabs in this software where an experimenter can click on. These tabs and sub tabs should be numbered to show the flow of procedures, otherwise the experimenter is not sure if he/she has completed all the necessary steps. The most important issue was uploading the data set, as this is the first step for using MSstatsQC software. If an experimenter cannot upload a data set with the correct format, he/she gets frustrated and may stop using the software. We can improve MSstatsQC by providing conspicuous guides close to the window related to uploading a data file as well as providing error messages that pop-up when the data set has a wrong format. Minor improvements include adding explanations to all the plots in the software, provide hints for interpreting the plots, spaced out graphs, bold plot titles, provide y-axis labels, and use the first 3 letters of peptide names to save the space. These minor improvements are important because the experimenter can spend less time to understand the plots and can interpret them more efficiently.


\bibliographystyle{unsrt}
%
\bibliography{main.bib}

\appendix

\section{Tasks}
In this section, we provide the exact tasks that we used Google forms.

We ask you to complete 4 tasks.  Estimated time to complete all the tasks is between 30 to 50 minutes. Each task is designed to measure usability of MSstatsQC or interpretability of the charts provided by the software. We will ask you to import data files and select appropriate metrics (task 1), interpret control charts (task 2), create decision rules (task 3) and interpret summary plots (task 4). 

To complete the tasks you need to have access to two files: data1.csv and data2.csv.  The files are in a google drive folder. Please download them by following this link:\\
https://drive.google.com/drive/folders/1jbR5DOv4uOae99K4-oU10Q4T3rLcXXDb?usp=sharing

Please let us know if you don't have access to the files. 

To begin, please use this link to open MSstatsQC software:\\
https://eralpdogu.shinyapps.io/msstatsqc/ 

Take your time to investigate the software before  clicking on next to start with task1:

\subsection{Task 1}
\textbf{Goal:}Data import, metric selection

\textbf{Estimated time to complete task 1:} 10 minutes\\

\textbf{Task 1:} Open data1.csv on your computer. Look at the data and it's columns. Make sure that the data has the right format based on the requirements of the software. Adjust the data if required. Select all the metrics available in data for further quality control analysis. Let the mean and standard deviation be selected from guide set. Do not change the default values of the guide set. Choose “CAVVDVPFGGAK” peptide for further quality control analysis. \\

About the guide set : The lower and upper bound of the guide set will determine what fraction of the samples in the data is used to estimate the mean and standard deviation. The default is 1 and 5, i.e., samples from row 1 to 5 is selected for estimating mean and standard deviation of each selected metric. \\

After completing the task, please answer the following questions:\\
Q1) Did you find any problem in the format of the data1.csv? If yes, how did you correct that.\\
Q2) Did you have any difficulty completing this task? If yes, please explain why.\\
Q3) Do you have any suggestions for the data import and metric selection part of the software?

\subsection{Task 2}
\textbf{Goal: }Interpret control charts

\textbf{Estimated time to complete task 1: } 10 minutes\\

\textbf{Task 2:} Clear the existing data table. Upload data2.csv to MSstatsQC shiny app. Make sure that the data has the right format based on the requirements of the software. Adjust the data if required. Select all the metrics available in data for further quality control analysis. Let the mean and standard deviation be selected from guide set. Do not change the default values of the guide set. Choose “CAVVDVPFGGAK” peptide for further quality control analysis. Go to “control charts” tab and select “XmR control charts”. Look into the sub-tabs (BestRetentionTime, MaxFWHM, TotalArea, Peak asymmetry) and familiarize yourself with the plots. Now, answer the following questions (Remember to talk loud with yourself”):\\
Q1) Do you understand the difference between the left and right plot? If not, please state the issue with the design.\\
Q2) For “CAVVDVPFGGAK” peptide, what time points shows the system is  out of control for Best Retention Time metric? Select all that apply.\\
5\\
6\\
7\\
8\\
9\\
12\\
13\\
19\\
20\\
21\\
22\\
23\\
24\\
26\\
29\\
30\\
Q3) For “CAVVDVPFGGAK” peptide, what time points shows the system is  out of control for MAX FWHM metric? Select all that apply.\\
5\\
6\\
7\\
8\\
9\\
12\\
13\\
19\\
20\\
21\\
22\\
23\\
24\\
26\\
29\\
30\\
Q4) Is there a time point that the system is out of control for at least 2 out of 4 available metrics? If yes, at what time point?\\
Q5) Do you have any suggestions about the system usability in data import and options tab, and interpretability of control charts?

\subsection{Task 3}
\textbf{Goal:}Create decision rules

\textbf{Estimated time to complete task 1:} 5 minutes\\

If you have already completed  task 2 do not read the rest of this paragraph. For this task, the CPTAC Study 9.1 data is in the datatask2 folder on the Desktop of this computer. Clear the existing data table. Upload CPTAC Study 9.1 data from the datatask2 folder on the Desktop to MSstatsQC shiny app. Make sure that the data has the right format. Adjust the data if needed.Select all the metrics available in data for further quality control analysis. Select all the metrics available in data for further quality control analysis. Let the mean and standard deviation be selected from guide set. Do not change the default values of the guide set. Choose “CAVVDVPFGGAK” peptide for further quality control analysis.

Select appropriate decision rules for the “Red flag” and  “Yellow flag” sections. Talk loud with yourself or write down your thoughts and explain why you selected these values. Remember that there is no right and wrong answer for this step.

\subsection{Task 4, part 1/3}
\textbf{Goal:}Interpret box plots

\textbf{Estimated time to complete task 1:} 5 minutes\\

Step 1: If you have already completed  task 2 or 3, do not read the rest of step 1 and start from step 2. Clear the existing data table. Upload data2.csv to MSstatsQC shiny app. Make sure that the data has the right format based on the requirements of the software. Adjust the data if required. Select all the metrics available in data for further quality control analysis. Select all the metrics available in data for further quality control analysis. Let the mean and standard deviation be selected from guide set. Do not change the default values of the guide set. Choose “CAVVDVPFGGAK” peptide for further quality control analysis.

Step 2: Go to “Create decision rules” tab. In the “Red flag” section, select 70 for “

Step 3: Go to “Metric summary” tab. Choose “Descriptives: boxplots for metrics” sub-tab. Answer the following questions (Remember to talk loud with yourself”):\\
Q1) Do you understand what each boxplot represents? \\
a)Yes\\     b) No\\
Q2) Do you think that providing this visualization for showing the range of values for each peptide in each metric is useful?\\
a)Yes\\     b) No\\
Q3) Do you think that setting different values for decision rules in step 4 will change the results in boxplots? Please justify your answer..\\
a)Yes\\     b) No\\
Q4) Do you have any suggestions for the design or usability of boxplots?\\

\subsection{Task 4, part 2/3}
\textbf{Goal:}Interpret decision maps

\textbf{Estimated time to complete task 1:} 5 minutes\\
Step 4: Now choose “Overall performance: decision maps” sub-tab. Let the control chart stay at “XmR chart”. Look at the decision maps and answer the following questions. (Remember to talk loud with yourself”):\\
Q1) Do you understand what each square in the decision map represent and what is the difference between the colors?\\
a) Yes\\     b) No\\
Q2) Do you understand the difference between the top and bottom decision map?\\
a) Yes\\     b) No \\
Q3) In top figure, Which metrics’ system performance is acceptable across all the runs? Select all that apply.\\
a) Total area\\     b) Peak assymetry     c) Max FWHM     d) Best retention time\\
Q4) In bottom figure, which metrics’ system performance is unacceptable at a certain run? Which time point? Select all that apply.\\
     a) Total area - time 37 \\    b) Peak assymetry - time 35 \\    c) Max FWHM - time 3 \\                           .    d) Best retention time - time 46 \\ e) Total area - time 3  \\   f) Peak assymetry - time 36    \\          .    g) Max FWHM - time 35 \\   h) Best retention time - time 37\\

Q5) Do you have any suggestions for the design of the decision maps?

\subsection{Task 4, part 3/3}
\textbf{Goal:}Interpret summary plots

\textbf{Estimated time to complete task 1:} 5 minutes\\
Step 5: Choose “Detailed performance: plot summaries” sub-tab. Let the control chart stay at “XmR chart”. Look at the “River plots” and “Radar plots” and answer the following questions. (Remember to talk loud with yourself”):\\
Q1) In River plots, at what time point the percentage of out of control peptides has the highest BestRetentionTime increase?\\
a) Between 0 - 10   \\   b) Between 10 - 20  \\   c) Between 20 - 30 \\    d) Between 30 - 40 \\
Q2) In Radar plots, which peptides has MAXFWHM  variability decrease? Select 4 that has the largest variability decrease. \\
CAVVDVPFGGAK\\
DDGSWEVIEGYR     \\
DGGIDPLVR \\
FFVAPFPEVFGK\\
GFCGLSQPK    \\
HLVDEPQNLIK \\
LVNELTEFAK \\
SLHTLFGDELCK\\
TAAYVNAIEK\\
VGPLLACLLGR\\
VLDALDSIK\\
VLVLDTDYK\\
YNGVFQECCQAEDK\\
YNLGLDLR\\
YSTDVSVDEVK\\
Q3) How hard was it to find the peptide names? (Scale 1-5)
1     2     3     4     5

\subsection{Questions about usability and interpretability of MSstatsQC}
\textbf{Estimated time to complete task 1:} 5 minutes\\
Q1) How easy was it to find what you are looking for on our website? (Convenience)\\
1     2     3     4     5\\
Q2) How visually appealing is our website? (Aesthetics)\\
1     2     3     4     5\\
Q3) How easy is it to understand the information on our website? (Comprehension)\\
1     2     3     4     5\\
Q4) How likely is it that you would recommend our website to a friend or colleague? (Satisfaction)\\
1     2     3     4     5\\
Q5) Is there something you found frustrating about the website? If yes, what was it?\\
Q6) Do you have any suggestions in terms of the design (fonts, colors, …) and usability (whether you achieve what you were looking for)\\

\end{document}